\documentclass[%
 reprint,
 amsmath,amssymb,
 aps,
 prl,
]{revtex4-2}
\usepackage{multirow}
\usepackage{aas_macros}
\usepackage{graphicx}
\usepackage{dcolumn}
\usepackage{bm}
\usepackage{amsmath}
\usepackage{caption} 
\usepackage{subcaption}
\usepackage{xcolor}

\newcommand{\fig}[1]{Fig.~\ref{#1}}

\newcommand{\dedf}{\ensuremath{\frac{\mathrm{d}E_{\mathrm{GW}}}}{\mathrm{d}f}}

\begin{document}

\preprint{APS/123-QED}

\title{The Stochastic Gravitational-Wave Background from Stellar Core-Collapse Events}

\author{Bella Finkel}

\affiliation{%
Skidmore College, Saratoga Springs, New York 12866, USA
}%

\author{Haakon Andresen}
\affiliation{%
The Oskar Klein Centre, Department of Astronomy, AlbaNova, SE-106 91 Stockholm, Sweden
}%

\author{Vuk Mandic}
\affiliation{%
School of Physics and Astronomy, University of Minnesota, Minneapolis, Minnesota 55455, USA
}

\date{\today}
\begin{abstract}
We estimate the stochastic gravitational-wave background arising from all stellar core-collapse events in the universe based on the gravitational-wave signal predictions of recent numerical simulations. We focus on waveforms from slowly or non-rotating stars and include rapidly rotating, highly massive progenitors as extreme case limits. Our most realistic estimates are more than one hundred times below the sensitivity of third-generation terrestrial gravitational-wave detectors and likely weaker than cosmological contributions to the stochastic gravitational-wave background.

\end{abstract}
\maketitle

\section{Introduction} \label{Introduction}
The stochastic gravitational-wave background (SGWB) arises 
from the superposition of gravitational-waves (GWs) from a variety of independent cosmological
\cite{DV2,Siemens,Auclair2020,Ringeval2017,Sousa2020,Blanco-Pillado2017,LIGOCosmicStrings,grishchuk,starobinskii,turner,Easther2007,Cook2012,Hogan1986,Witten1984,Kosowsky1992,Romero2021,Lopez2015,Geller2018,caprini,Buonanno1997,GASPERINI2003,Mandic2006,Gasperini2016}
and astrophysical sources 
\cite{LIGOCBC2017,LIGOBBHC2016,Cholis2017,Farmer2003,Regimbau2006,Cheng2017,Buonanno2005,Marassi2009,Crocker,Sandick2006,mandicPBBH}. We focus on the SGWB due to stellar core-collapse events \cite{Buonanno2005,Marassi2009,Crocker,Sandick2006}. 
At the end of their lifetime, stars larger than about $8\ \mathrm{M_\odot}$ form a gravitationally unstable iron core which collapses until its inner layer
overshoots nuclear density and a proto-neutron star (PNS) is formed. The still
collapsing outer core ``bounces'' off the inner region and launches an outwards propagating shock wave.
The shock loses energy as it passes through the core and eventually stalls. 
In most cases, shock revival is thought to be powered by the absorption of a fraction of the neutrinos emitted from the PNS by material behind the shock, which causes the shock to break out the stellar surface
and reveals the explosion as an electromagnetic transient.

In a normal, non-rotating supernova
most of the GW emission is expected to come from oscillations of the PNS, but
other hydrodynamic processes can contribute a large fraction of the emission. Low-frequency emission (below $300\ \mathrm{Hz}$) mainly comes from the standing accretion shock instability (SASI), a hydrodynamic instability in the gain region that drives large scale spiral and sloshing oscillations of the shock and is expected to arise in a fraction of core-collapses \cite{Blondin2003}.
High-frequency emission (above $300\ \mathrm{Hz}$) largely arises from oscillations of the PNS.

The non-detection of GW transients associated with core-collapse events has constrained the GW energy produced by core-collapse supernovae (CCSNe) and ruled out regions of the parameter spaces of extreme emission models \cite{LIGO2016,LIGO2020}. However, these constraints on GW emission remain a few orders of magnitude higher than the predictions of multidimensional numerical simulations. Additionally, the first three observing runs of Advanced LIGO and Advanced Virgo and the first two data releases by the International Pulsar Timing Array have provided upper limit estimates for the isotropic SGWB \cite{LIGO2021upper,Verbiest2016,Antoniadis2022}.

Previous studies \cite{Coward2002,Buonanno2005,Crocker} computed the SGWB due to CCSNe using, predominantly, 
two-dimensional simulations. Two-dimensional simulations are unable to capture the effects of non-axisymmetric instabilities and tend to systematically overestimate GW amplitudes \cite{Andresen2017}.
We sample a variety of theoretical predictions of the GW signal from 
three-dimensional numerical core-collapse simulations with sophisticated hydrodynamics and neutrino transport treatments, of duration from hundreds of milliseconds to upwards of a second.
We begin by reviewing the procedure for estimating the background of astrophysical sources. Next, we give an overview of the details of the numerical simulations and their GW signals. Finally, we discuss features of the background and its detectability by
third-generation GW detectors, such as Cosmic Explorer (CE) \cite{CosmicExplorer}.

\section{Calculation of the SGWB from CCSNe}\label{sec:CalculationoftheSGWBfromCCSNe}
The stochastic gravitational-wave background is usually described by its dimensionless energy spectrum:
\begin{equation}\label{eqn:OmegaNorm}
    \Omega_{\mathrm{GW}}(f) = \frac{1}{\rho_c}\frac{\mathrm{d}\rho_{\mathrm{GW}}}{\mathrm{d}\ln f},
\end{equation}
where $\rho_{GW}$ is the GW energy density in the frequency band $(f,f+df)$ and $\rho_c$ is the critical energy density which gives a flat universe, $\rho_c=\frac{3H_0^2c^2}{8\pi G}.$ Here, $c$ is the speed of light, $G$ is Newton's gravitational constant, and $H_0$ is the Hubble constant ($=67.7\mathrm{\ km/s/Mpc}$ \cite{LIGO2021upper}). 

The normalized GW energy density $\Omega_\mathrm{GW}$ can be expressed in terms of the energy spectrum emitted by a single CCSN, $\frac{\mathrm{d}E_{\mathrm{GW}}}{\mathrm{d}f_e}(f_e)$, by the equation
\begin{equation}\label{eqn:OmegaSFR}
    \Omega_{\mathrm{GW}}(f)=\frac{f}{\rho_c H_0}\int_0^{z_{\mathrm{max}}}\mathrm{d}z\frac{R(z)\frac{\mathrm{d}E_{\mathrm{GW}}}{\mathrm{d}f_e}(f_e)}{(1+z)E(\Omega_m,\Omega_\Lambda,z)},
\end{equation} 
where $f_e$ is the frequency emitted at the source ($f_e=f(1+z)$).  
$R(z)$ is the rate of stellar core-collapse events per comoving volume as a function of redshift, assumed to follow the star formation rate (SFR) $R_*(z)$ as 
\begin{equation}\label{eqn:SCCR}
    R(z) = \lambda_{\mathrm{CC}}R_*(z),
\end{equation}
where $\lambda_{\mathrm{CC}}$ is the mass fraction of stars which experience core collapse. We estimate $\lambda_{\mathrm{CC}}$ from the Salpeter initial mass function (IMF) $\phi(m)=Nm^{-2.35}$ with the normalization $\int^{\infty}_{0.1\mathrm{M}_\odot}{\phi(m)m\mathrm{d}m=1}$
and the assumption that all stars whose mass is greater than $8 \; \mathrm{M}_\odot$ undergo collapse. ($\mathrm{M_\odot}$ denotes units of solar mass.) With these assumptions, 
\begin{equation}
    \lambda_{\mathrm{CC}}=\int^{\infty}_{8\mathrm{M}_\odot}{\phi(m)\mathrm{d}m}\approx0.007 \; \mathrm{M^{-1}_\odot}.
\end{equation}
While this $\lambda_{\mathrm{CC}}$ estimate is approximate, we do not expect its uncertainty to be larger than a factor of two, implying that the uncertainty in $\lambda_{\mathrm{CC}}$ will not make a qualitative difference in our results. 

We choose the SFR model proposed by \cite{Vangioni2015}. This model 
fits the parametrized form from Springel \& Hernquist \cite{Hernquist2003},
\begin{equation}
    R_*(z)=\nu\frac{pe^{q(z-z_m)}}{p-q+qe^{p(z-z_m)}},
\end{equation}
to the galaxy SFR derived by \cite{Behroozi2013} to obtain the parameter values $\nu=0.178\mathrm{\ M_\odot/yr/Mpc^3},$ $z_m=2.00,$ $p=2.37,$ and $q=1.80$. This model is based on observations of the galaxy luminosity function at high redshift. Using the SFR given in Eq. (15) of \cite{Madau2014} based on UV and IR data gives only marginally lower background estimates which do not differ from our results by more than 30\%. As discussed below, variations in the spectra obtained from simulations constitute a significantly larger source of uncertainty than the particular SFR chosen.

Eq. (\ref{eqn:OmegaSFR}) includes a factor of $1+z$ to account for cosmic expansion and convert time from the source frame to  the observation frame. The function 
\begin{equation}\label{eqn:zComVol}
    E(\Omega_m,\Omega_\Lambda,z) =  \sqrt{\Omega_m(1+z)^3+\Omega_\Lambda}
\end{equation}
expresses the comoving volume's dependence on redshift. $\Omega_m$ and $\Omega_\Lambda$ are the energy density in matter and in dark energy for a flat cosmological model, $\Omega_m=0.311$ and $\Omega_\Lambda=0.689$ \cite{Planck2020}. Writing Eq. (\ref{eqn:OmegaSFR}) in terms of the SFR in Eq. (\ref{eqn:SCCR}), we have

\begin{equation}\label{eqn:Crocker5}
    \Omega_{\mathrm{GW}}(f)=\frac{8\pi Gf\lambda_{\mathrm{CC}}}{3H^3_0 c^2}\int \mathrm{d}z \frac{R_*(z)\frac{\mathrm{d}E_{\mathrm{GW}}}{\mathrm{d}f_e}(f_e)}{(1+z)E(\Omega_m,\Omega_\Lambda,z)}.
\end{equation}
Note that Eq. (\ref{eqn:Crocker5}). does not account for the delay time between the formation of a star and its core collapse.
We have investigated the effect of this delay and found it to be negligible.

The spectral energy density of the GWs emitted by a core-collapse supernova is required to calculate its contribution to $\Omega_{\mathrm{GW}}$. The energy radiated as GWs to infinity by a source is given by \cite{gravity}
\begin{equation}
    E_{\mathrm{GW}}=\int_0^t\mathrm{d}\tau\int T_{0\nu} n^\nu r^2 \mathrm{d}\Omega,
\end{equation}
where the angular integral should be performed over a spherical shell at infinity that encloses the source. The vector $n^\nu$ is a spacelike unit vector perpendicular to the surface of the spherical shell, $r$ denotes the radial coordinate of a spherical coordinate system, $t$ is the time duration of the signal, and $T_{\mu\nu}$ denotes the GW energy-momentum tensor which, in the transverse-traceless (TT) gauge, is given by
\begin{equation}
    T_{\mu\nu}=\frac{c^5}{32\pi G}\big\langle(\partial_\mu h^{\mathrm{TT}}_{\gamma\lambda})(\partial_\nu h^{\gamma\lambda}_{\mathrm{TT}})\big\rangle,
\end{equation}
here $\langle...\rangle$ denotes averaging over several wavelengths. Using  the symmetries of the  energy-momentum tensor in the TT-gauge we obtain

\begin{equation}\label{eqn:EnergyRadiated}
    E_\mathrm{GW}=\frac{c^3}{16\pi G}\int_0^t\mathrm{d}\tau\int r^2\big\langle(\partial_t h_\times^{\mathrm{TT}})^2+(\partial_t h_{+}^\mathrm{TT})^2\big \rangle\mathrm{d}\Omega
\end{equation}
(see \cite{gravity} and \citep{EMuller2012} for details).

By applying Parseval’s theorem to Eq. (\ref{eqn:EnergyRadiated}) one can show that the spectral energy density of the GW emission is given by

\begin{equation} \label{eq:dedf}
    \dedf = \frac{c^3}{16\pi G} (2\pi f)^2 \int r^2 
    \big \langle (\tilde{h}_{\times}^{\mathrm{TT}})^2+ (\tilde{h}_{+}^\mathrm{TT})^2\big \rangle
    \mathrm{d}\Omega,
\end{equation}
where $\tilde{h}_\times^\mathrm{TT}$ and  $\tilde{h}_+^\mathrm{TT}$ denote the Fourier transforms of the cross and
plus polarization modes of the GWs in the TT gauge, respectively \citep{EMuller2012,andresen_thesis}. Note that Eq. (\ref{eq:dedf}) will be slightly modified for a discrete  signal and will depend on the exact normalization and implementation of the discrete Fourier transform used, see \cite{andresen_thesis} for details.

The angular dependence of the GW signal is needed for Eq. (\ref{eq:dedf}), but most numerical simulations
do not provide this data. It is customary to publish the GW  signal emitted in a few selected directions, which means that we are required to approximate the angular integral based on the data available to us.

\section{Core-Collapse Simulations}
We have begun to see convergences in the predictions of
CCSNe GW signals
\citep{EMuller2012,Yakunin2015,Andresen2015,Kuroda2016,Andresen2017,OConnor2018,Andresen2019,
Powell2019,Radice2019,VartanyanBurrowsRadice2019,Powell2020,Mezzacappa2020,Andresen2021,Powell2021}, 
but the details of GWs emitted by CCSNe are stochastic 
and exhibit a large degree of variation. 
To ensure that our results are robust
and represent the variation inherent to supernova GWs, we include a large set of theoretical signal predictions
from numerical simulations. We have selected waveforms produced by different groups, with
different numerical codes, and with different input physics.
We focus on supernovae from non- or slowly-rotating progenitors, 
which should make up more than 90 percent of all CCSNe.
\begin{table*}[]
\begin{tabular}{cccccc}
Model name & ZAMS mass, type & Numerical code  & EoS & Notes & Reference \\
\hline \hline
        m39   &   39 $\mathrm{M_\odot}$, Wolf-Rayet star & \multirow{6}{*}{\textsc{CoCoNut-FMT} \cite{MullerJanka2015}}     & LS220 & Rotating, Exploding &  \multirow{3}{*}{\cite{Powell2020}}\\
        s18np &  18 $\mathrm{M_\odot}$, giant &    & LS220 &  SASI \\ 
        y20   &   20 $\mathrm{M_\odot}$, Wolf-Rayet star &    &  LS220 & Exploding & \\
        s18   &   18 $\mathrm{M_\odot}$, giant &     &  LS220 & Exploding  & \cite{Powell2019} \\
        z100 & 100 $\mathrm{M_\odot}$ & & SFHx & SASI & \multirow{3}{*}{\cite{Powell2021}} \\
        z85 & 85 $\mathrm{M_\odot}$ & &  SFHx & Exploding, SASI &  \\
        \hline
        Rad9  &   $9\ \mathrm{M_\odot}$ &  \multirow{8}{*}{\textsc{Fornax} \cite{Skinner2019}}    & SFHo & Exploding & \multirow{8}{*}{\cite{Radice2019}} \\
        Rad10  &   $10\ \mathrm{M_\odot}$ &   & SFHo & Exploding &  \\
        Rad11  &   $11\ \mathrm{M_\odot}$ &    & SFHo & Exploding&  \\
        Rad12  &   $12\ \mathrm{M_\odot}$ &     & SFHo & Exploding&  \\
        Rad13  &   $13\ \mathrm{M_\odot}$ &     & SFHo & &  \\
        Rad19  &   $19\ \mathrm{M_\odot}$ &   & SFHo & Exploding &  \\
        Rad25  &   $25\ \mathrm{M_\odot}$ &   & SFHo & Exploding, SASI &  \\
        Rad60  &   $60\ \mathrm{M_\odot}$ &   & SFHo & Exploding &  \\
        \hline
        s9-FMD-H   &   9 $\mathrm{M_\odot}$, giant & \multirow{2}{*}{\textsc{Aenus-Alcar} \cite{Obergaulinger2015,Just2015}}     & SFHo & Exploding  & \multirow{2}{*}{\cite{Andresen2021}} \\
        s20-FMD-H  &   20 $\mathrm{M_\odot}$, giant &     &
        SFHo &  \\
        \hline
        s15nr & 15 $\mathrm{M_\odot}$ & \multirow{3}{*}{\textsc{Prometheus-Vertex} \cite{Rampp2002}}  &LS220 & SASI & \multirow{3}{*}{\cite{Andresen2019}}\\
        s15r & 15 $\mathrm{M_\odot}$ &  &LS220 &  SASI&  \\
        s15fr & 15 $\mathrm{M_\odot}$ &  & LS220 & Rotating, Exploding, SASI& \\
        \hline
        mesa20-pert  &   20 $\mathrm{M_\odot}$, giant &  \multirow{2}{*}{\textsc{FLASH} \cite{Weaver2017}}  & SFHo & SASI & \multirow{2}{*}{\cite{OConnor2018}} \\
        mesa20 &   20 $\mathrm{M_\odot}$, giant &   & SFHo & SASI &  \\
        \hline
        Shib0 & 70 $\mathrm{M_\odot}$ & \multirow{3}{*}{\cite{KurodaCode2016}}  & LS220 & SASI &\multirow{3}{*}{\cite{Shibagaki2021}} \\
        Shib1 & 70 $\mathrm{M_\odot}$ &  & LS220 &  Rotating, low-$T/|W|$ instability & \\
        Shib2& 70 $\mathrm{M_\odot}$ &  & LS220 & Rotating, low-$T/|W|$ instability & \\

\end{tabular}
\caption{Simulations from which we calculate the SGWB. The high-density nuclear equations of state (EoS) include SFHo \& SFHx \cite{Steiner2013} and that of Lattimer \& Swesty \cite{LATTIMER1991} with bulk incompressibility of $\mathrm{K=220 \ MeV}$ (LS220). }
\label{tab:sims}
\end{table*}

\begin{figure*}[ht]
    \centering
    \includegraphics[width=\linewidth]{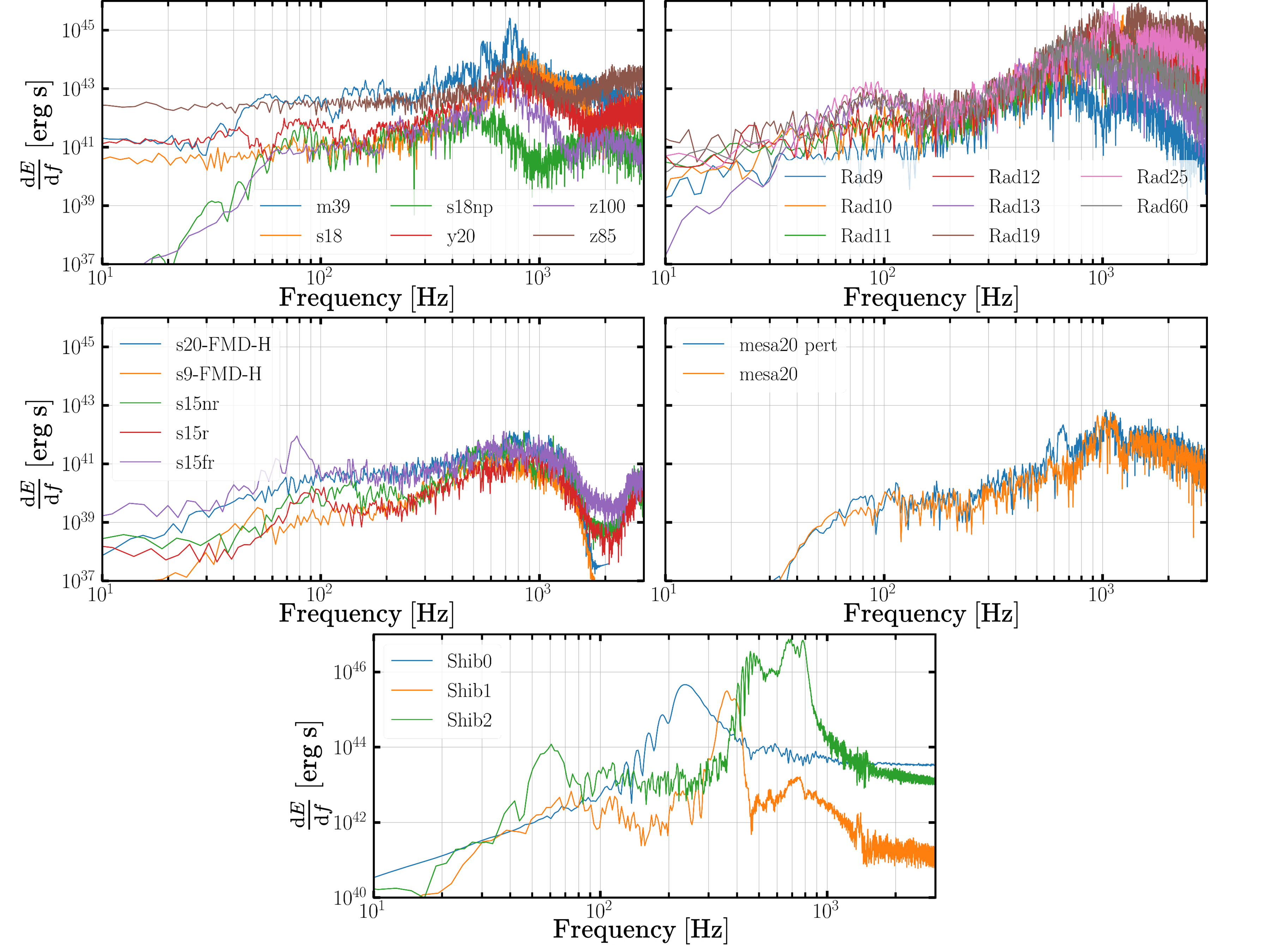}
    \caption{Total time-integrated GW spectra $\frac{\mathrm{d}E}{\mathrm{d}f}$ for emission from models by Powell \& M\"uller \cite{Powell2019,Powell2020,Powell2021} (upper left), Radice et al. \cite{Radice2019} (upper right), Andresen et al. \cite{Andresen2019,Andresen2021} (middle left), O'Connor \& Couch \cite{OConnor2018} (middle right), and Shibagaki et al. \cite{Shibagaki2021} (bottom). The vertical axis range on the plot showing the models by Shibagaki et al. is higher than that for the other panels.}
    \label{fig:Spectra}
\end{figure*}

Table \ref{tab:sims} summarizes details of the simulations we use to calculate the background. We denote the models of \cite{Radice2019} by ``Rad'' followed by the mass of the progenitor, and those of \cite{Shibagaki2021} by ``Shib'' followed by their rotation rate. Otherwise, we name the models as they are presented in the original papers.

We choose simulations from \cite{Powell2019,Powell2020} which model core collapse in two non-rotating
$18 \ \mathrm{M_\odot}$ red supergiants (s18 and s18np), a rapidly rotating $39 \ \mathrm{M_\odot}$ Wolf-Rayet star (m39), and a non-rotating $20 \ \mathrm{M_\odot}$ Wolf-Rayet star (y20). Model s18 is seeded with density perturbations from convective oxygen burning in its oxygen shell to aid in its explosion. No such perturbations are included in the s18np simulation, which prevents shock revival and allows the development of significant SASI activity. 
Model m39 has an initial surface rotation velocity of $600 \ \mathrm{km \; s^{-1}}$. Its high mass and rapid rotation produce strong GW emission, particularly at the equator.

Powell, M\"uller \& Heger model two non-rotating Pop-III stars of masses $100 \ \mathrm{M_\odot}$ (z100) and $85 \ \mathrm{M_\odot}$ (z85) in the pulsational pair instability regime \cite{Powell2021}. Model z85 undergoes shock revival and collapses to a black hole before the end of its simulation, whereas the simulation of model z100 is terminated before shock revival or black hole collapse. Both models demonstrate strong GW emission across the frequency range we examine.

Radice et al. \cite{Radice2019}  model the collapse of seven non-rotating progenitors with ZAMS masses of $9\ \mathrm{M_\odot},$ $10\ \mathrm{M_\odot},$ $11\ \mathrm{M_\odot},$ $12\ \mathrm{M_\odot},$ $13\ \mathrm{M_\odot},$ $19\ \mathrm{M_\odot},$ $25\ \mathrm{M_\odot},$ and $60\ \mathrm{M_\odot}$.
For all of these simulations, PNS oscillations are the main GW source. The models emit low-frequency GWs caused by prompt convection, then develop neutrino-driven convection or, in the case of Rad25, the SASI. 
The shock revives in all models except Rad13. Apart from Rad9, all exploding models show asymmetric accretion onto the PNS at late times and significant continued emission at the termination of the simulation. 

From Andresen et al. \cite{Andresen2021} we select the two models s9-FMD-H and s20-FMD-H. The $9\ \mathrm{M_\odot}$ model, s9-FMD-H, exhibits a low accretion rate onto the PNS and primarily emits at frequencies above $300 \ \mathrm{Hz}$ with some low-frequency GW emission. The $20\ \mathrm{M_\odot}$ model,
s20-FMD-H, develops hot-bubble convection and significant SASI activity. 
The SASI activity appears as GW emission at frequencies below $250\ \mathrm{Hz}$.

Andresen et al. \cite{Andresen2019} investigate the effects of moderate rotation with three $15\ \mathrm{M_\odot}$ progenitors: one non-rotating (s15nr), one with central angular velocity $\Omega_0=0.2\ \mathrm{rad\ s^{-1}}$ (s15r), and one fast rotating (s15fr) with central angular velocity $\Omega_0=0.5\ \mathrm{rad\ s^{-1}}$. We include all three modes in our analysis.
Model s15fr experiences strong spiral SASI activity in the post-shock flow before explosion. The SASI spiral mode is strong in s15nr. SASI activity is significantly weaker in s15r than in the other models, and the postshock region is dominated by convection. 

O'Connor and Couch simulate the collapse of a $20\ \mathrm{M_\odot}$ zero-age main-sequence star \cite{OConnor2018}.
We use the GW signal from their three-dimensional, standard resolution models mesa20 and mesa20\_pert, the latter of which is seeded with velocity perturbations in the silicon and oxygen shells when the simulation is mapped to \textsc{FLASH} \cite{Weaver2017}. Both mesa20 and mesa20\_pert demonstrate SASI activity. Neither of these models explodes.

We also consider the $70\ \mathrm{M_\odot}$ progenitors simulated with central angular velocities of $\Omega_0=0\ \mathrm{rad\ s^{-1}}$, $\Omega_0=1\ \mathrm{rad\ s^{-1}}$, and $\Omega_0=2\ \mathrm{rad\ s^{-1}}$ (Shib0, Shib1, and Shib2) by \cite{Shibagaki2021}. Before collapse, this star has a central iron core mass of $\sim4.6\ \mathrm{M_\odot}$.
When the simulations are terminated, Shib0, Shib1, and Shib2 have PNS masses of $\sim2.5\ M_\odot$, $\sim2.2\ M_\odot$, and $\sim2.6\ \mathrm{M_\odot}$, respectively. The sloshing and spiral SASI modes of Shib0 produce strong GW emission
between $200$ and $300\ \mathrm{Hz}$.
The low-$T/\vert W\vert$ instability, where $T/\vert W\vert$ refers to the ratio of rotational and gravitational potential energy, is a rotational instability that can produce strong GW emission. It is observed in both rotating models, resulting in emission from $400$ to $800 \ \mathrm{Hz}$ in Shib2 and from $300$ to $400 \ \mathrm{Hz}$ in Shib1. 

Fewer than ten percent of CCSNe are expected to arise from rapidly rotating progenitors \cite{Janka2012}. Thus, it should be noted that the GW background spectra calculated below using models m39 and Shib2 represent the unrealistic scenario where all stars which undergo core collapse are rapidly rotating.

\section{Results}

\begin{figure*}[ht]
    \centering
    \includegraphics[width=\linewidth]{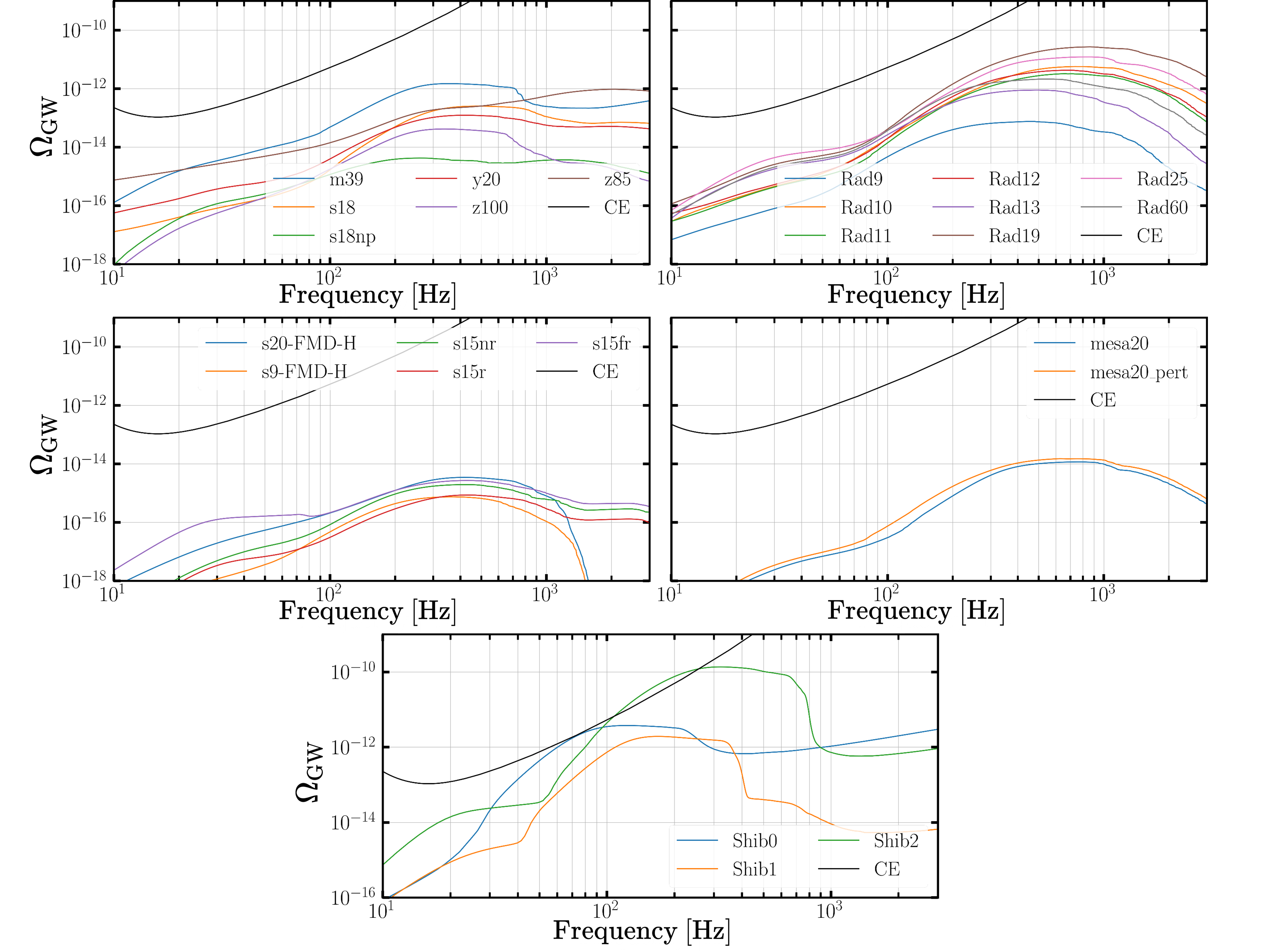}
    \caption{$\Omega_{\mathrm{GW}}$ for models by Powell \& M\"uller \cite{Powell2019,Powell2020,Powell2021} (upper left), Radice et al. \cite{Radice2019} (upper right), Andresen et al. \cite{Andresen2019,Andresen2021} (middle left), O'Connor \& Couch \cite{OConnor2018} (middle right) and Shibagaki et al. \cite{Shibagaki2021} (bottom) shown in comparison with the $2\sigma$ power-law integrated sensitivity \cite{PhysRevD.88.124032} of two collocated Cosmic Explorer \cite{CosmicExplorer} detectors assuming one year exposure. Note that these spectra are calculated under the assumption that all CCSNe events emit the same GW spectrum. In particular, the Shib1 and Shib2 models shown in the bottom panel are calculated under the unrealistic assumption that all CCSNe progenitors rotate rapidly and exhibit the low-$T/\vert W\vert$ instability.}
    \label{fig:Omega}
\end{figure*}

\begin{figure}[ht]
    \centering
    \includegraphics[width=\linewidth]{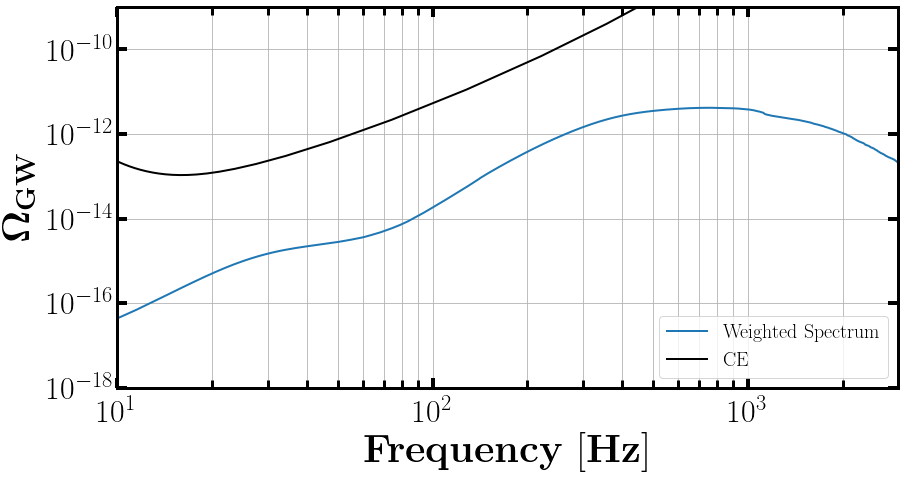}
    \caption{Averaged $\Omega_{\mathrm{GW}}$ including the contributions of the non-rotating progenitors and excluding Shib0 weighted by the abundance of the stellar progenitor in the stellar population as given by the Salpeter IMF (c.f. Eq. \ref{Eq:sum_omega}). 
    }
    \label{fig:WeightedSpectrum}
\end{figure}

We use the core-collapse GW energy spectra $\mathrm{d}E/\mathrm{d}f$ shown in \fig{fig:Spectra} to calculate the normalized GW energy density $\Omega_{\rm GW}(f)$. Our results are shown in Fig. \ref{fig:Omega}. For each curve in 
Fig. \ref{fig:Omega} we assume that all CCSNe have the same GW emission spectrum given by the corresponding core-collapse simulation model. As discussed above, the spectral properties of the signal are related to the physical properties of the supernova core (see \cite{Pajkos2021,Warren2020,Torres-Forne2019} for further details).

All models yield backgrounds that would be undetectable by second-generation GW detectors, whose sensitivity is expected to reach $\Omega_{\rm GW}(f) \sim 10^{-9}$ - $10^{-10}$. Most models yield backgrounds at least two orders of magnitude below the sensitivity of third-generation GW detectors, under the assumption of cross-correlating two collocated detectors of CE sensitivity \cite{CosmicExplorer} with one year of exposure. 

The exceptions are the models by \cite{Shibagaki2021}, which become borderline observable by CE, see the bottom panel of Fig. \ref{fig:Omega}. However, this conclusion assumes that all CCSNe progenitors are rapidly rotating and develop the energetically-emitting low-$T/\vert W\vert$ instability. When these models are weighted by their prominence in the stellar population (about $1\%$ of CCSNe are expected to arise from rapidly rotating progenitors \cite{Janka2012}), their contribution to the overall SGWB becomes comparable to that of the other models. 
Considering only non- and slowly rotating models, Rad25 and Rad19 provide the strongest SGWB, and the model s15r yields the weakest SGWB.

We note that our results are subject to the uncertainty in the rate of CCSNe events across the universe, modelled in Eq. (\ref{eqn:Crocker5}) by the star formation rate $R_*$ and the scaling factor $\lambda_{\rm CC}$. While the scaling factor may be uncertain at the level of a factor of two, the star formation rate model is rather robust especially at redshifts smaller than two, which is where the majority of the CCSNe SGWB comes from. 

We provide an averaged estimate of the background based on the relative abundances of non-rotating progenitors in Fig. \ref{fig:WeightedSpectrum}. The contribution of each model to the background is weighted by its prominence in the stellar population as given by the normalized Salpeter IMF. We define the average spectrum 
\begin{equation}
    \Omega_A(f)=\frac{1}{\Gamma}\sum_{i}\phi(m_i)\Omega_{i}(f)\Delta m_i,
    \label{Eq:sum_omega}
\end{equation}
where $\Omega_{i}(f)$ is the background spectrum for the model(s) of mass $m_i$, and $\Delta m_i$ extends from the halfway point between $m_i$ and the model of the next lowest mass to the halfway point between $m_i$ and the model of the next highest mass. In cases where there is more than one model of the same mass, we use a simple average between their background spectra for $\Omega(m_i)$. In Eq. (\ref{Eq:sum_omega}), the normalization constant $\Gamma$ is given by $\Gamma=\int_{8\mathrm{M_\odot}}^{100\mathrm{M_\odot}}\phi(m)\mathrm{d}m$. It should be noted that the absence of a one-to-one relationship between the mass of the progenitor and the GW signal of its collapse, the stochastic nature of the signal, and the limited selection of core-collapse models mean that Fig. \ref{fig:WeightedSpectrum} is only of illustrative value, and should not be taken as a decisive estimate of the stochastic gravitational-wave background produced by
core-collapse supernovae.


\section{Conclusions}
In this work, we have estimated a range for the SGWB resulting from stellar core-collapse events using the results of three-dimensional numerical simulations of a variety of progenitors. The GWs emitted during core collapse are stochastic and associated with multiple emission processes. We find that in all but the most extreme cases, the SGWB from CCSNe is $2\mbox{-}5$ orders of magnitude below the sensitivity of the third-generation GW detectors.

Our results are limited by the approximations used in the simulations we draw from and the characteristics of the stellar population we sample.
All of the simulations shown in this paper were terminated at different times, 
in most cases while still emitting significant gravitational radiation. In extreme core-collapse scenarios appreciable GW emission occurs late in the explosion phase \cite{Jardine2021,Raynaud2021}. Longer simulations would provide a fuller assessment of the GW signal and a better picture of the background.

Anisotropic neutrino emission from the PNS can cause GW emission in the range of sub-Hertz to hundreds of Hertz \cite{Epstein1978,EMuller1997,Kotake2009,Vartanyan2020,Richardson2021,Mukhopadhyay2021}. A more complete description of the background would include the contribution of this signal, in particular at frequencies below the lower $10\ \mathrm{Hz}$ limit of our background calculations. 
Asymmetries resulting from the presence of magnetic fields can also lead to GW emission; magnetohydrodynamic effects have been found to significantly alter the GW signal \cite{Raynaud2021,Muller2020MHD}.

At $10\ \mathrm{Hz}$, the lower limit of the frequency range considered here,
the core-collapse contribution to the SGWB is $\Omega_{\rm GW}(f) \sim10^{-15}$ or weaker, meaning that cosmological GW backgrounds may be detectable over the core-collapse GW background.
In many cases, the cosmological SGWB models are a few orders of magnitude stronger than our highest estimates for the CCSN background.
For example, the backgrounds from cosmic string networks in scaling with string tension $G\mu$
greater than $10^{-17}$ \cite{Auclair2020,Sousa2020}, primordial binary black hole coalescences involving masses greater than $10^{-6}\ \mathrm{M_\odot}$ \cite{Wang2019,Mukherjee2021},
and inflation mechanisms (e.g., those with an effective graviton mass between $m=0\ \mathrm{H}$ and $m=0.9\ \mathrm{H}$, with a Hubble rate during inflation of $H=10^{12}\ \mathrm{GeV}$ or $H=10^{13}\ \mathrm{GeV}$ and tensor sound speed between $c_T=0.2$ and $c_T=1$, in natural units \cite{Bartolo2016}) are expected to be one or more orders of magnitude above our predictions for the core-collapse background.
These sources may therefore be accessible to future detectors, and their visibility will likely be unimpeded by the background due to CCSNe.

\begin{acknowledgments}
VM would like to acknowledge the support of the NSF grant PHY-2110238. 
BF would like to acknowledge the support of the NSF REU grant 2049645. 
HA is supported by the Swedish Research Council (Project No. 2020-00452). The paper has been assigned designator LIGO-P2100353.
\end{acknowledgments}


\bibliography{citations}

\end{document}